\documentclass[11pt]{article}
\usepackage{hyperref}
\pdfoutput=1

\begin{document}
\title{Vortex Induced Oscillations of Cylinders}
\author{Roberto Camassa, Bong Jae Chung, Greg Gipson\\
 Richard M. McLaughlin, Ashwin Vaidya 
\\ \vspace{6pt} Department of Mathematics, \\ University of North Carolina, Chapel Hill, NC 27599, USA}
\maketitle

\begin{abstract}
 The submitted fluid dynamics videos \cite{viovideo} depict the various orientational dynamics of a hinged cylinder suspended in a flow tank. The different behaviors displayed by the cylinder range from steady orientation to periodic oscillation and even autorotation. We illustrate these features
using a phase diagram which captures the observed phenomena as a function of Reynolds number ($Re$) and reduced inertia ($I^*$). A hydrogen bubble flow visualization technique is also used to show vortex shedding structure in
the cylinder's wake which results in these oscillations.
\end{abstract}

\section{Introduction}

The question of orientation bodies in fluids dates back to Kirchoff\cite{kirchoff} who examined the
dynamics of falling paper. The steady state orientation of bodies with certain classes of
symmetries has been long known. A sedimenting cylinder, for instance, is known to fall with
its axis of symmetry perpendicular to gravity in a Newtonian fluid when its length exceeds
its diameter. However, in the case of a disk, when the length of the cylinder is less than
the diameter, the disk falls with its axis of rotation parallel to gravity. Several more recent
systematic studies have experimentally, theoretically and numerically explored the steady
state dynamics of falling bodies (see \cite{galdi,hu} for instance and references therein) in the Stokes and low inertial
regimes. The orientational dynamics becomes even more interesting at higher Reynolds numbers as 
vortex shedding effects become significant and give rise to oscillations of the body.

\section{Our experiments}

The (\href{http://ecommons.library.cornell.edu/handle/1813/11484/vio(large).mpg}{ video})  \cite{viovideo} shows experiments on the orientational dynamics of a hinged cylinder immersed in a flow.  The experimental setup consists of plastic cylinders of diameter 0.635 cm and lengths
ranging from 0.32cms-1.27cms. The aspect ratio,(length/diameter), ranged from 0.5 - 2.0. The
cylinders were made of ABS, Lexan and Delrin (plastics) with densities 1.05g/cc, 1.18g/cc
and 1.4g/cc, respectively and were placed at the center of the water tunnel and 
suspended there by means of a taut stainless steel wire of thickness 0.023cms passing through a hole of
diameter 0.04cms at the center of the particle. The particles were suspended in such a way as to
allow it to oscillate freely only along an axis, perpendicular to the flow direction. The dynamics of the cylinder were recorded using a 72mm SONY HDR FX1
camera fitted with a zoom lens. The operational speed of the camera
was set at 30 frames per second and the resulting videos were analyzed using the Video Spot
Tracker V.5.20 program\cite{spot}.\\

We note several transitions in the behavior of the cylinder, depending
upon the inertia of the fluid and the body. These include (i) steady state orientation, (ii) random oscillations, (iii) periodic oscillations and (iv) autorotation. We plot a phase diagram which characterizes the different behaviors displayed by the cylinder as a function of Reynolds number, $Re=Ul/\nu $ and reduced inertia, $I^*=I/\rho_f d^5 $ \cite{paolo}, where $U$ is the centerline velocity of the fluid in the absence of any obstacle, $I$ is the moment of inertia of the cylinder with respect to its axis of revolution, $l$ is the maximum of length or diameter of the cylinder, $d$ is the diameter of the cylinder, $\rho_f$ is the density of the fluid and $\nu$ its kinematic viscosity. The $Re$ achieved in this experiment ranges from about 100-6000 while
the values of $I^*$ typically range from 0.1-0.6. The video shows some of the phenomena indicated above and also some hydrogen bubble flow visualization of the vortex shedding in the wake, primarily for cylinders with length/diameter $ \approx 1$, which are the most dynamic particles. Please note that the frame rate for some of the movies with hydrogen bubble flow visualization have been reduced by a factor of 2 for visual effect. We refer the readers to \cite{paolo} for more information about this work and for further relevant references on this subject.\\

\end{document}